\documentclass{article}
\usepackage{spconf}
\usepackage{graphicx}
\usepackage{amsmath,amsfonts,amssymb,amsthm}
\usepackage{enumitem}
\usepackage{verbatim}
\usepackage{pgfplots}
\usepackage{tikz}
\usepackage{color}
\usetikzlibrary{arrows,shapes,chains,matrix,positioning,scopes,patterns,calc,
decorations.markings,
decorations.pathmorphing,
}
\usepackage{hyperref}

\def\Ktot{K_\mathrm{tot}}
\def\Ka{K_\mathrm{a}}

\newtheorem{theorem}{Theorem}
\newtheorem{lemma}[theorem]{Lemma}

\newtheorem{remark}[theorem]{Remark}

\newcommand{\av}{\vec{\mathrm{a}}}
\newcommand{\bv}{\vec{\mathrm{b}}}
\newcommand{\xv}{\vec{\mathrm{x}}}
\newcommand{\yv}{\vec{\mathrm{y}}}
\newcommand{\zv}{\vec{\mathrm{z}}}

\newcommand{\defeq}{\triangleq}

\definecolor{mycolor1}{rgb}{0.63529,0.07843,0.18431}%
\definecolor{mycolor2}{rgb}{0.00000,0.44706,0.74118}%
\definecolor{mycolor3}{rgb}{0.00000,0.49804,0.00000}%
\definecolor{mycolor4}{rgb}{0.87059,0.49020,0.00000}%
\definecolor{mycolor5}{rgb}{0.00000,0.44700,0.74100}%
\definecolor{mycolor6}{rgb}{0.74902,0.00000,0.74902}%

\newif\ifproof
\prooffalse

\title {A Coupled Compressive Sensing Scheme for Unsourced Multiple Access}
\name{\begin{tabular}{c}Vamsi K. Amalladinne,
Avinash Vem,
Dileep Kumar Soma, \\
	Krishna R. Narayanan,
Jean-Francois Chamberland \end{tabular}
\thanks{
This material is based upon work supported by the National Science Foundation (NSF) under Grant No.~CCF-1619085.}
}
\address{Department of Electrical and Computer Engineering, Texas A\&M University}
\begin{document}

\maketitle

\begin{abstract}
This article introduces a novel paradigm for the unsourced multiple-access communication problem.
This divide-and-conquer approach leverages recent advances in compressive sensing and forward error correction to produce a computationally efficient algorithm.
Within the proposed framework, every active device first partitions its data into several sub-blocks, and subsequently adds redundancy using a systematic linear block code.
Compressive sensing techniques are then employed to recover sub-blocks, and the original messages are obtained by connecting pieces together using a low-complexity tree-based algorithm.
Numerical results suggest that the proposed scheme outperforms other existing practical coding schemes.
Measured performance lies approximately $4.3$~dB away from the Polyanskiy achievability limit, which is obtained in the absence of complexity constraints.
\end{abstract}


\section{Introduction}
\label{sec:intro}

Unsourced multiple access communication (MAC), initially proposed by Polyanskiy~\cite{polyanskiy2017perspective}, is a novel formulation for concurrent uplink data transfers.
It is closely related to uncoordinated multiple access~\cite{paolini2015coded,chen2017capacity}.
In this new paradigm, a system contains a total of $\Ktot$ users; out of this group, $\Ka$ users each wish to transmit a $B$-bit message to the access point at any given time.
The access point is tasked with recovering only the set of messages being transmitted, without regard for the identities of the corresponding sources.
The total number of users $\Ktot$ can be very large, whereas parameters $\Ka$ and $B$ are envisioned to remain small, typically in the hundreds.

For the regime of interest, with its characteristic short message lengths, non-asymptotic information-theoretic results apply.
Along these lines, Polyanskiy~\cite{polyanskiy2017perspective} derives finite block-length, achievability bounds for the unsourced MAC.
The findings reported therein are based on random Gaussian codebooks, and they assume that information is recovered using a maximum likelihood decoder.
Algorithmically, this scheme can be very computationally demanding.
In~\cite{ordentlich2017low}, Ordentlich and Polyanskiy report that many existing multiple access strategies perform poorly in this context, especially when $\Ka$ exceeds $100$.
They also propose the first low-complexity coding scheme tailored to this setting.
In their scheme, a transmission period is divided into sub-blocks, or slots, and the system operates in a synchronous fashion.
That is, all the users are aware of slot boundaries.
Within this framework, every active user transmits a codeword during a randomly chosen slot.
A data block is formed with a concatenated code that is designed for a $T$-user real-addition Gaussian multiple access channel ($T$-GMAC); typical values for $T$ range from 2 to 5.
Although this proposed scheme performs significantly better than existing MAC protocols, there remains an important gap of approximately $20$~dB between its performance and the achievability limit associated with the unsourced MAC~\cite{polyanskiy2017perspective}.
In related work~\cite{vem2017user}, we introduce a low-complexity coding scheme that relies on a similar slotted structure.
Our previous framework consists of an improved, close-to-optimal coding strategy for the $T$-GMAC; coupled to the application of successive interference cancellation across slots to reduce the performance degradation caused by overcrowded slots.
The combination of these two features constitutes a significant improvement over previous results~\cite{ordentlich2017low}, with a performance curve that lies only approximately $6$~dB away from the above mentioned achievability limit.

Both schemes discussed above adopt a channel coding viewpoint wherein the $\Ka$-user GMAC is reduced to multiple smaller $T$-GMAC channel problems.
Contrastingly, in this paper, we develop an alternate compressive sensing (CS) view of the problem.
To begin, we emphasize that a naive CS solution to the unsourced MAC entails resolving a $2^B$-length linear problem with a $\Ka$-sparse solution.
This implies sensing matrices with $2^{100}$ columns, which renders the problem intractable.
A key idea in establishing a pragmatic scheme is to divide the information blocks of the users into smaller sub-blocks such that each sub-block is amenable to a CS recovery.
Before transmission, redundancy is added to individual sub-blocks using a systematic linear block code.
The collection of sub-blocks transmitted within a slot are recovered using a CS algorithm.
Once this is achieved, individual segments of the original messages need to be pieced together.
This is accomplished via a low-complexity tree-based algorithm.
The overall structure of this communication architecture yields better performance compared to other existing algorithms with comparable computational complexity.

Throughout, we employ $\mathbb{R}_{+}, \mathbb{Z}_{+}$, and $\mathbb{N}$ to denote the non-negative real numbers, non-negative integers, and natural numbers, respectively.
For any $a,b \in  \mathbb{Z}_{+}$ with $a \le b$, we use $[a:b]$ to denote $\{c \in \mathbb{Z}_{+}: a \le c \le b \}$.
We write $X \sim B(n,p)$ if a random variable $X$ possesses a binomial distribution with parameters $n$ and $p$.
We employ $|A|$ for the cardinality of set $A$, and we use $[x]$ to denotes the closest integer to $x$.

\section{SYSTEM MODEL}
\label{sec:systemmodel}

Let $\mathbf{S}_\mathrm{tot}$ represent the collection of devices within a network, and let $\mathbf{S}_\mathrm{a}$ denote the subset of active devices within a communication round, $\mathbf{S}_\mathrm{a} \subset \mathbf{S}_\mathrm{tot}$.
Then, we have $|\mathbf{S}_\mathrm{tot}| = \Ktot$ and $|\mathbf{S}_\mathrm{a}| = \Ka$.
Every active device wishes to communicate $B$ bits of information to a base station through an uncoordinated uplink transmission scheme.
The number of channel uses dedicated to this process is $N$, and $\mathit{W} = \left\{ \vec{\mathit{w}}_k:k \in \mathbf{S}_\mathrm{a} \right\}$ represent the collection of $B$ bit message vectors associated with these active devices.
We assume that devices pick their message vectors independently and uniformly at random from the set of binary sequences $\{0, 1\}^B$.

The base station facilitates a slotted structure for multiple access on the uplink through coarse synchronization.
As such, the signal available at the receiver assumes the form
\[
\textstyle
\yv= \sum_{k \in \mathbf{S}_\mathrm{a}} \xv_k + \zv,
\]
where $\xv_k$ is the $N$-dimensional vector sent by device~$k$ and $\zv$ represents additive white Gaussian noise.
The signal sent by every device is power constrained, i.e., $\| \xv_k \|_2^2 \le NP$ for $k \in \mathbf{S}_\mathrm{a}$, a scenario akin to \cite{polyanskiy2017perspective}.
The energy-per-bit is then given by $\frac{E_b}{N_0} \triangleq \frac{NP}{2B}$. The receiver produces an estimate $\widehat{\mathit{W}}(\yv)$ for the list of transmitted binary vectors $\mathit{W}$ with $| \widehat{\mathit{W}}(\yv) | \le K_\text{a}$.
The per-user error probability of the system is defined by
\begin{equation}
P_\mathrm{e}
= \textstyle
\frac{1}{K_\text{a}}\sum_{k \in \mathbf{S}_\mathrm{a}}
\Pr \left( \vec{\mathit{w}}_k \notin \widehat{\mathit{W}}(\yv) \right) .
\end{equation}
We propose an encoding and decoding scheme that achieves $P_\mathrm{e} \le \varepsilon$, where $\varepsilon$ is the target error probability with manageable computational complexity.

\section{PROPOSED SCHEME}
\label{sec:proposedscheme}

\begin{figure}[!ht]
\centering
\resizebox{\columnwidth}{!}{\begin{tikzpicture}
\def\nodewidth{0.6in}
\def\fsize{\Large}
\def\sfsize{\normalsize}
\def\xoffs{0.5in}
\def\ya{2.5}
\def\xr{4.5}
\def\xadder{9}
\def\xslots{13.5}
\def\xdec{20}
\tikzstyle{block} = [rectangle, draw, thick, opacity=0.7,line width =2, minimum size=\nodewidth]
\tikzstyle{vertRectangle} = [rectangle, draw, opacity=0.7,line width =2, minimum width=\nodewidth, minimum height=8*\nodewidth]
\tikzstyle{opnode} = [rectangle, draw, thick,opacity=0.7,line width=1, minimum size=0.2in]

\foreach \i/\j in {1/1,2/i,4/\Ka}{ 
\node[block] (r1\i) at (0,-\i*\ya){\fsize Tree Encoder};
\node[block] (r2\i) at (\xr,-\i*\ya) {\fsize CS Encoder};

\draw[<-, thick, line width=2] (r1\i.west)--node[midway, above]{\fsize $\vec{\mathit{w}}_{\j}$}+(-\xoffs,0);
\draw[->, thick, line width=2] (r1\i.east)--node[midway, above]{\fsize $\tilde{\vec{\mathit{w}}}_{\j}$}(r2\i.west);
}
\node  at (0.5*\xr,-3*\ya) {\Huge $\vdots$};

\node[draw,circle] (adder) at (\xadder,-2.5*\ya) {\Large $\sum$};

\draw[thick] (r21.east) -- +(1.8,0) [->]-- (adder.north);
\draw[thick] (r22.east) -- +(1.6,0) [->]-- (adder.north west);
\draw[thick] (r24.east) -- +(1.4,0) [->]-- (adder.south west);
\draw[thick,<-] (adder.south) -- node[midway,right] {\Large $\vec{\mathrm{z}}$} +(0,-\xoffs);

\node[vertRectangle,align=center] (r3) at (\xslots,-2.5*\ya){\fsize \bf CS decoder};
\node[vertRectangle,align=center] (r4) at (\xdec,-2.5*\ya){\fsize \bf Tree Decoder};

\draw[<-, thick] (r3.north west)++(0,-0.5*\nodewidth)-- +(-1,0) -- (adder.north east);
\draw[<-, thick] (r3.north west)++(0,-1.5*\nodewidth)-- +(-0.7,0) --(adder.north east);
\draw[<-, thick] (r3.south west)++(0,0.5*\nodewidth)-- +(-0.9,0) --(adder.south east);

\draw[transform canvas={yshift=-\nodewidth},thick] (r3.north west) -- (r3.north east);
\draw[transform canvas={yshift=-2*\nodewidth},thick] (r3.north west) -- (r3.north east);
\draw[transform canvas={yshift=\nodewidth}] (r3.south west) -- (r3.south east);
\node [below=0.3*\nodewidth of r3.north](s1) {\fsize  sub-block 0};
\node [below=1.3*\nodewidth of r3.north](s2) {\fsize sub-block 1};
\node [above=0.3*\nodewidth of r3.south](s4) {\fsize sub-block n-1};
\node [below=3*\nodewidth of r3.north]() {\Huge $\vdots$};
\node [above=2*\nodewidth of r3.south]() {\Huge $\vdots$};

\draw[transform canvas={yshift=-\nodewidth},thick] (r4.north west) -- (r4.north east);
\draw[transform canvas={yshift=-2*\nodewidth},thick] (r4.north west) -- (r4.north east);
\draw[transform canvas={yshift=\nodewidth}] (r4.south west) -- (r4.south east);
\node [below=0.3*\nodewidth of r4.north]() {\fsize  Depth 0};
\node [below=1.3*\nodewidth of r4.north]() {\fsize Depth 1};
\node [above=0.3*\nodewidth of r4.south]() {\fsize Depth n-1};
\node [below=3*\nodewidth of r4.north]() {\rotatebox{-90}{\Huge $\rightarrow$}};
\node [above=2*\nodewidth of r4.south]() {\rotatebox{-90}{\Huge $\rightarrow$}};

\begin{scope}[very thick,decoration={
    markings,
    mark=at position 0.5 with {\arrow{>}}}
    ] 
\draw[transform canvas={yshift=-0.5*\nodewidth},postaction={decorate}] (r3.north east) -- (r4.north west);
\draw[transform canvas={yshift=-1.5*\nodewidth},postaction={decorate}] (r3.north east) -- (r4.north west);
\draw[transform canvas={yshift=-2.5*\nodewidth},postaction={decorate}] (r3.north east) -- (r4.north west);
\draw[transform canvas={yshift=0.5*\nodewidth},postaction={decorate}] (r3.south east) -- (r4.south west);
\end{scope}

\draw[->, thick, line width=2] (r4.east)--node[midway, above]{\fsize $\widehat{\vec{\mathit{w}}}_1,\ldots, \widehat{\vec{\mathit{w}}}_{K_a}$}+(2.5*\xoffs,0);
\end{tikzpicture}}
\caption{This is a schematic of the proposed scheme.
Original messages are split into sub-blocks, and redundancy is added to individual components.
Transmitted sub-signals are then determined via a CS matrix, and sent over a MAC channel.
A CS decoder recovers lists of sub-blocks, and a tree decoder reconstructs the orginal messages.}
\label{fig:1}
\end{figure}
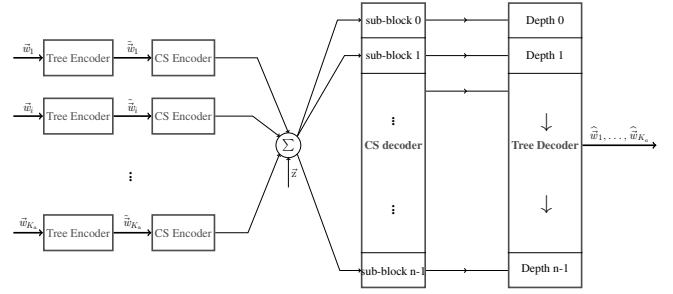

%
The transmission strategy outlined above features two parts: a systematic linear block code based on random parity checks, which we refer to as the tree encoder, and a CS encoder.
A notional diagram of the proposed system appears in Fig.~\ref{fig:1}.

\textbf{\textit{Tree Encoder:}}
Every $B$-bit binary message vector $\vec{\mathit{w}}$ is encoded into $M$ bits using a systematic linear block code, which has random parity check constraints.  Algorithmically, a message vector is partitioned into $n$ sub-blocks, with the $i^{\text{th}}$ sub-block consisting of $m_i$ message bits, $\sum_{i=0}^{n-1}m_i=B$.
The tree encoder appends $l_i$ parity bits to sub-block~$i$, except for the first block as we choose $l_0 = 0$.
All the coded sub-blocks have the same length, i.e., $m_i+l_i =J\defeq  {M}/{n}$. 
The parity check bits in each sub-block are constructed as follows.
Let $\big( p^{(i)}_{0},p^{(i)}_{1},\dots,p^{(i)}_{l_{i}-1} \big)$ denote the parity bits in sub-block $i$.
These bits are selected to satisfy random parity check constraints for all the message bits preceding their respective sub-block.
To this end, we concatenate the message bits of all the sub-blocks $k \in [0:i]$ and index them with the set $\big[ 0:\sum_{k=0}^{i}m_k-1 \big]$. We then choose $l_{i}$ subsets $\mathcal{A}^{(i)}_j \subseteq \big[ 0:\sum_{k=0}^{i}m_k-1 \big]~\forall ~j \in [0:l_i-1]$ uniformly at random without replacement.
Parity check $p^{(i)}_j$ is chosen as the modulo-2 sum of all the message bits indexed by the set $\mathcal{A}^{(i)}_j$.
In effect, $p^{(i)}_j$ acts as a parity check constraint for some randomly chosen message bits preceding it.
In Section~\ref{parityvector}, we describe an optimization framework for the choice of parity length vector $\vec{l}=(l_0=0,l_1.\dots,l_{n-1})$.

\textbf{\textit{CS Encoder:}} Let $\mathbf{A}=[\av_1,\ldots,\av_{2^J}] \in \{\pm \sqrt{P}\}^{\tilde{N}\times2^J}$, where $\tilde{N}\defeq {N}/{n}$, denote a compressed sensing matrix that is designed to recover any $\Ka$-sparse binary vector in the presence of noise with a low probability of error.  The $J$ bits in a sub-block are encoded  using a bijective function $f:\{0,1\}^J \rightarrow \{\av_j, j \in [1:2^J]\},$ which maps each sub-block to a column in $\mathbf{A}$.
That is, a column of $\mathbf{A}$ is a potentially transmitted sub-block.



\subsection{Decoder}
\label{decoder}
The decoding scheme consists of two components: the CS decoder operating in each sub-block, and a tree decoder operating across sub-blocks.

\textbf{\textit{CS Decoder:}} The signal received during the $i^{\text{th}}$ sub-block can be expressed as $\yv_i = \mathbf{A}\bv_i + \zv_i$,
where $\bv_i \in \{0,1\}^{2^J}$ is a $\Ka$-sparse binary vector that indicates the list of $i^{\text{th}}$ sub-blocks transmitted by the active users.
The task of the CS decoder is to provide an estimate of the sparse vector $\bv_i$ from the received signal $\yv_i$ during the corresponding time slot.
This is accomplished by first applying a non-negative least squares (NNLS) algorithm to get an estimate $\vec{b}^{(\text{nnls})}_i$ of vector $\vec{b}_i$.
Yet, this does not ensure that the entries of vector $\vec{b}^{(\text{nnls})}_i$ are binary.
The desired binary estimate $\hat{\vec{b}}_i$ is obtained by setting the $K$ largest entries of the vector $\bv^{(\text{nnls})}_i$ to one and the remaining $2^J-K$ entries to zero.
The number $K$ is chosen as $K=\Ka+K_{\delta}$, where $K_{\delta}$ is a small positive integer.
Although the list output by the CS decoder is larger than $\Ka$, the quantity $K_{\delta}$ is carefully chosen such that the erroneously decoded sub-blocks are very unlikely to satisfy the parity check constraints associated with encoding process.

\textbf{\textit{Tree Decoder:}} The tree decoder seeks to recover the original messages transmitted by all the users by piecing together valid sequences of elements drawn from the various CS lists.
Towards this end, the access point constructs a decoding tree for \textit{each candidate message} as follows.
We fix a sub-block from the list of all possible first sub-blocks supplied by the CS decoder as the root node for a tree.
Once the first sub-block is determined, there are $K$ possible choices for the second sub-block, and these are the nodes which appear in the first stage of the tree.
Similarly, there are $K$ possible choices for the third sub-block for each choice of the second sub-block and, hence, $K^2$ nodes in the second stage.
This process continues until the $(n-1)^{\text{th}}$ stage is reached; at this point, the tree has $K^{n-1}$ leafs.
Every path connecting the root node to a leaf becomes a possible message.
If there exist a single valid path at the end, the decoder outputs the corresponding message; otherwise, it reports a failure.

The number of possible paths increases exponentially with the stages of the tree and, hence, a naive search through all the leaf nodes is infeasible.
In practice, invalid paths are pruned iteratively through the parity check constraints.
Specifically, at stage $i \ge 1$, the decoder retains only nodes that satisfy the $l_i$ bit parity constraints on all the message bits preceding that stage.
This iterative procedure continues until the $(n-1)^{\text{th}}$ stage is reached.
The complexity of this decoding scheme depends on the number of nodes surviving each stage, since parity checks have to be enforced only on the children of surviving nodes in the subsequence stages of the tree decoding process.

\begin{remark}[Iterative Extension]
\label{rmk:iterations}
The successful outputs from the tree decoder can be subtracted off from their respective received signals in each sub-block.
This extra step can potentially improve the estimate provided by the CS decoder compared to the previous iteration.
This successive interference cancellation method can be repeated iteratively, leading to significant potential gains in performance, particularly for the first few steps.
\end{remark}

\section{Performance Analysis}
\label{perr}
Suppose that the list output by the CS decoder contains the sub-block $i$ transmitted by user $k$ with a probability $1-p_{cs}$ and, with a probability $p_{cs}$, this block is erroneously replaced by a vector chosen uniformly at random from the set $\{0,1\}^{J}$.
Let $E_k$ denote the event that the transmitted binary message form user~$k$ is not present on the list output by the tree decoder.
Similarly, let $C_k$ be the event that all the sub-blocks corresponding to this user are present on the lists output by the CS decoder.
Probability $P(E_k)$ can be computed as,
\begin{align}
P(E_k) = P(E_k\lvert C_k)P(C_k) + P(E_k\lvert \overline{C_k})P(\overline{C_k}).  \label{Totalprob}
\end{align}
If the CS decoder fails to decode at least one of the sub-blocks that correspond to a user, then the output of the tree decoder would not contain the original message transmitted by that user.
Thus, we have $P(E_k \lvert \overline{C_k})=1$.
The quantity $P(C_k)$ can be computed as $P(C_k)=(1-p_{cs})^n$.
We denote the event that the tree decoder declares a failure because of more than one path surviving the tree decoding process by $E_k\lvert C_k$.
We write the probability of this event as $p_{\mathrm{tree}}$.
When there are no iterations involved in the decoding process, the quantity $P_e$ is the same as $P(E_k)$; they can be computed using \eqref{Totalprob} and the above observations as $P_e = 1-(1-p_{\mathrm{tree}})(1-p_{cs})^n$.


Let $L_i$ denote the random variable corresponding to the number of erroneous paths that survive stage $i \in[1:n-1]$ of the tree decoding process.
The following results hold.
\begin{lemma} \label{Lemma1}
Expected values for $L_i$ are given by
\begin{align} \label{expLi}
\mathbf{E}[L_{i}] &= \sum_{m=1}^{i} \left( K^{i-m}(K-1)
\textstyle \prod_{j=m}^{i}p_j \right)
\end{align}
where $p_i ={1}/{2^{l_i}}$, $q_i= 1-p_i$, and $i \in [1:n-1]$.
\end{lemma}
\begin{lemma} \label{Lemma2}
The probability of error for the tree decoder $p_{\mathrm{tree}}$ is given by
$p_{\mathrm{tree}} = 1-G_{L_{n-1}}(0)$ where
\begin{align}
G_{L_{n-1}}(z) &= \textstyle \prod_{i=0}^{n-2} f_{n-1-i}^{K-1}(z) \notag \\
f_k(z) &= \begin{cases}
    q_k + p_k f_{k+1}^{K}(z),& ~1 \le k \le n-1 \\
    z^{\frac{1}{K}},              & ~k=n,
\end{cases} \label{ptree}
\end{align}
and $p_i, q_i$ are given in \eqref{expLi}.
\end{lemma}

\begin{table*}[!tbh]
\centering
 \begin{tabular}{||c|| c| c| c| c| c| c| c| c| c| c| c| c||}
 \hline
$\Ka$ & 25 & 50 & 75 & 100 & 125 & 150 & 175 & 200 & 225 & 250 & 275 & 300 \\[0.4ex]
 \hline
 $J$ & 14 & 14 & 14 & 14 & 14 & 15 & 15 & 15 & 15 & 15 & 15 & 15 \\[0.4ex]
 \hline
$\varepsilon_{\text{tree}}$ &0.0025 & 0.0045 & 0.006 & 0.01 & 0.0125 & 0.0055 & 0.0065 & 0.007 & 0.008 & 0.01 & 0.0125 & 0.0175 \\[0.4ex]
 \hline
\end{tabular}
\caption{Various parameters used in simulations.}
 \label{tableofparams}
\end{table*}

We define the computational complexity $C$ of this decoder as the number of nodes on which parity checks need to be performed.
\begin{lemma}
\label{Lemma3}
A closed-form expression for computing the expected computational complexity $\mathbf{E}[C]$ is given by
\begin{equation*} 
\mathbf{E}[C] = K \left( n-1
+ \sum_{i=1}^{n-2} \sum_{m=1}^{i} \left( K^{i-m}(K-1)
\textstyle \prod_{j=m}^{i} p_j \right) \right)
\end{equation*}
where $p_i, q_i$ are given in \eqref{expLi}.
\end{lemma}
The proof of Lemma~\ref{Lemma1} relies on the fact that $L_i \lvert L_{i-1} \sim B((L_{i-1}+1)K-1,p_i)$.
The proof of Lemma~\ref{Lemma2} is based on computing a closed-form expression for the probability generating function (PGF) $G_{L_{n-1}}(z)$ of the random variable $L_{n-1}$.
Lemma~\ref{Lemma3} is a straightforward extension of Lemma~\ref{Lemma1}.
Additional details about these proofs can be found in~\cite{vamsi2017tree}.

\subsection{Choice of the Parity Length Vector}
\label{parityvector}
We formulate the constrained optimization problem of minimizing the expected complexity subject to the probability of decoding failure being less than a carefully chosen threshold $\varepsilon_{\mathrm{tree}}$.
Since the parity lengths are non-negative integers, such a problem would be very difficult to solve.
As such, we relax the problem to $(l_1,l_2,\dots,l_{n-1}) \in \mathbb{R}^{n-1}_+$. Also, we replace the constraint $p_{\mathrm{tree}} \le \varepsilon_{\mathrm{tree}}$ with $ \mathbf{E}\left[L_{n-1}\right] \le \varepsilon_{\mathrm{tree}}$ for the purpose of mathematical tractability.
(By Markov's inequality, the quantity $\mathbf{E}\left[L_{n-1}\right]$ is an upper bound on $p_{\mathrm{tree}}$).
After these modifications, the optimization framework for the choice of parity lengths is given by
\begin{equation}
\begin{aligned}
& \underset{(p_1, p_2 , \dots, p_{n-1})}{\text{minimize}}
& &\mathbf{E}[C] \\
& \text{subject to}
& & \mathbf{E}\left[L_{n-1}\right] \le \varepsilon_{\mathrm{tree}}, \\
&&& \textstyle \sum_{i=1}^{n-1} \log_2\left({1}/{p_i}\right) = M-B, \\
&&& p_i \in \left[{1}/{2^{J}},1\right]~ \forall~i \in [1:n-1]. \label{optprob3}
\end{aligned}
\end{equation}
The above is a geometric program \cite{boyd2004convex}, and it can be solved using standard convex solvers.
We choose the parity check lengths as $\hat{l}_i = \left[ \log_2 \left({1}/{\hat{p}_i}\right) \right]$, for $i \in [1:n-1]$ where $\big( \hat{p}_1, \hat{p}_2, \dots \hat{p}_{n-1} \big)$ is the solution to the optimization problem.

\begin{figure}[!ht]
\centering
\resizebox{0.5\textwidth}{!}{\begin{tikzpicture}
\def\fsize{\large}
\pgfplotsset{every y tick label/.append style={font=\fsize}}
\pgfplotsset{every x tick label/.append style={font=\fsize}}

\begin{axis}[%
width=4in,
height=3in,
scale only axis,
every outer x axis line/.append style={white!15!black},
every x tick label/.append style={font=\color{white!15!black}},
xmin=25,
xmax=300,
xtick = {50,100,...,300},
xlabel={\fsize Number of active users $\Ka$},
xmajorgrids,
every outer y axis line/.append style={white!15!black},
every y tick label/.append style={font=\color{white!15!black}},
ymin=0,
ymax=25,
ytick = {0,5,...,25},
ylabel={\fsize Required $E_b/N_0$ (dB)},
ymajorgrids,
legend style={at={(0,1)},anchor=north west, draw=black,fill=white,legend cell align=left,font=\fsize}
]

\addplot [color=black,dotted,line width=2.0pt]
  table[row sep=crcr]{
 25	0.25\\
50	0.3\\
75	0.35\\
100	0.4\\
125	0.45\\
150	0.5\\
175	0.55\\
200	0.6\\
225	0.95\\
250	1.25\\
275	1.55\\
300	1.8\\
};
\addlegendentry{Random Coding\cite{polyanskiy2017perspective}};

\addplot [color=mycolor1,dotted,line width=2.0pt]
  table[row sep=crcr]{25	2.26\\
50	2.88\\
75	3.9\\
100	5.03\\
125	5.8798\\
150	7.3954\\
175	8.6199\\
200	9.7328\\
225	11.1761\\
250	12.6127\\
275	13.3907\\
300	14.9116\\
};
\addlegendentry{4-fold ALOHA\cite{ordentlich2017low}};
\addplot [color=red,solid,line width=2.0pt,mark size=3.5pt,mark=diamond,mark options={solid}]
  table[row sep=crcr]{25	3.9849\\
50	5.6396\\
75	6.2855\\
100	6.7952\\
125	7.5262\\
150	8.3122\\
175	9.1418\\
200	10.103\\
225	11.062\\
250	12.279\\
275	13.296\\
300	14.6648\\
};
\addlegendentry{SIC T=2\cite{vem2017user}};
\addplot [color=mycolor2,solid,line width=2.0pt,mark size=1.4pt,mark=square,mark options={solid}]
  table[row sep=crcr]{25	3.18\\
50	3.52\\
75	4.64\\
100	5.61\\
125	5.85\\
150	6.46\\
175	6.72\\
200	7.41\\
225	7.6772\\
250	8.3217\\
275	8.8428\\
300	9.352\\
};
\addlegendentry{SIC T=4\cite{vem2017user}};
\addplot [color=mycolor3,solid,line width=2.0pt,mark size=2.0pt,mark=+,mark options={solid}]
  table[row sep=crcr]{25	4.4\\
50	4.8\\
75	5.16\\
100	5.5\\
125	5.85\\
150	6.9\\
175	7.5\\
200	7.95\\
225	8.7\\
250	9.4167\\
275	10.2773\\
300	11.37\\
};
\addlegendentry{Proposed Scheme, 0 iterations};
\addplot [color=mycolor4,solid,line width=2.0pt,mark size=1.3pt,mark=triangle,mark options={solid,rotate=90}]
  table[row sep=crcr]{25	3.54\\
50	3.6\\
75	3.74\\
100	3.8\\
125	4.02\\
150	4.85\\
175	5.23\\
200	5.52\\
225	6.05\\
250	6.67\\
275	7.4\\
300	8.22\\
};
\addlegendentry{Proposed Scheme, 1 iteration};
\addplot [color=mycolor5,dotted,line width=2.0pt]
  table[row sep=crcr]{25	7.5\\
35	7.3\\
50	8.75\\
100	11.7\\
150	14.5\\
200	18\\
250	21\\
300	23\\
};
\addlegendentry{OP-Exact\cite{ordentlich2017low}};
\node[] at (axis cs: 200,4.8) {\scriptsize X};
\node[] at (axis cs: 200,4.43) {\scriptsize O};

%

\end{axis}

\end{tikzpicture}%
}
\caption{Minimum $E_b/N_0$ required to acheive $P_e \le 0.05$ vs.\ number of users for various schemes.
Results for $2$ and $3$ iterations (see Remark~\ref{rmk:iterations}) are represented by `x' and `o'.
Observe that the SNR gains diminish with each iteration.}
\label{fig:sim_results}
\end{figure}
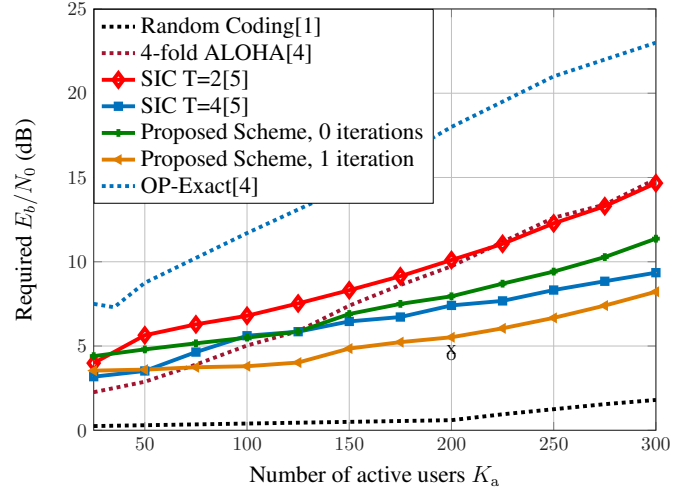

\section{Simulation Results}
\label{sec:simresults}
In this section, we study the performance of the proposed framework and we provide comparisions with existing schemes in literature.
We consider a system with $\Ka \in [25:300]$ active users, each having $B=75$~bits of information to transmit.
We divide these bits into $n=11$ sub-blocks and the quantity $J$, which denotes the length of each sub-block is chosen depending on $\Ka$;
it is given in Table~\ref{tableofparams}.
Similar to \cite{vem2017user}, we use sensing matrices that are constructed based on BCH codes for the compressed sensing problem. Specifically, we select a subset $\mathcal{C}^0$ of codewords of size $|\mathcal{C}^0|=2^J$ from the (2047,23) BCH codebook $\mathcal{C}$ with the following properties: \textbf{(i)} $\vec{c} \in \mathcal{C}^0 \implies \vec{1} \oplus \vec{c} \in \mathcal{C}\setminus \mathcal{C}^0$, where $\vec{1} \oplus \vec{c}$ denotes the one's complement of $\vec{c}$; \textbf{(ii)} $\vec{c}_1, \vec{c}_2 \in \mathcal{C}^0 \implies \vec{c}_1 + \vec{c}_2 \in \mathcal{C}^0$; \textbf{(iii)} $\vec{0} \in \mathcal{C}^0$ where $\vec{0}$ denotes the all zero codeword.
We then choose the sensing matrix as $\mathbf{A} = [\vec{a}_0, \vec{a}_1, \cdots, \vec{a}_{2^J-1}]$, with dimension $2047 \times 2^J$ where $\vec{a}_i=\sqrt{P}(2\vec{c}_i-1), \vec{c}_i \in \mathcal{C}^0$ for $i \in [0:2^J-1]$. The total number of channel uses is therefore given by $N=11\times2047=22,517$. The target error probability of the system is fixed at $\varepsilon=0.05$. We set list size $K$ for the NNLS CS problem to $K=\Ka+10$. For each $\Ka \in [25:300]$, we solve the optimization problem \eqref{optprob3} using the CVX solver~\cite{grant2008cvx}, and the resulting solution dictates the choice of parity length vector.
Choice of the quantity $\varepsilon_{\mathrm{tree}}$ for each $\Ka$ is given in Table~\ref{tableofparams}.
The parameters $B$ and $N$ are chosen such that the rate $\frac{B}{N}=\frac{75}{22,517}$ is approximately the same as the rate resulting from the choice of parameters $B=100$ and $N=30,000$ in \cite{ordentlich2017low,vem2017user}.
This enables a fair comparison between these schemes and our proposed scheme.
We emphasize that the choice of $B$ and $N$ for our simulations is motivated by the existence of good compressive sensing matrices based on BCH codes.
When these parameters are proportionally scaled up, performance of the system can only improve, as the finite block length effects are more pronounced for lower values of $B$ and $N$.
In Fig.~\ref{fig:sim_results}, the $E_b/N_0$ required to achieve a target error probability of 0.05 is plotted as a function of $\Ka$ for various schemes.
%
It can be seen from Fig.~\ref{fig:sim_results} that our proposed scheme with just one extended round of iteration outperforms existing schemes for $\Ka \in [75:300]$. 

\bibliographystyle{IEEEbib}
\bibliography{IEEEabrv,MACcollision}

\end{document}